\documentclass{ecai}
\usepackage{times}
\usepackage{graphicx}
\usepackage{latexsym}

\usepackage{amsthm}
\usepackage{color}  
\usepackage{dsfont}
\usepackage{mathtools}
\usepackage{amsmath}
\usepackage{amssymb}
\usepackage{algorithm}
\usepackage{algpseudocode}
\usepackage{multirow}
\usepackage{booktabs}
\usepackage{dblfloatfix}
\usepackage{subcaption}
\usepackage{url}

\newcommand{\Rr}{{\mathbb{R}}}

\newcommand{\Nn}{{\mathbb{N}}}

\newcommand{\by}{{\bf y}}

\newcommand{\lb}{\left(}
\newcommand{\rb}{\right)}

\ecaisubmission   

\begin{document}

\title{Explain Graph Neural Networks to Understand  Weighted Graph Features in Node Classification}

\author{Xiaoxiao Li \institute{Yale University, USA, email: xiaoxiao.li@yale.edu} \and Jo\~ao Sa\'ude \institute{J.P. Morgan AI Research,
USA, email: joao.saude@jpmorgan.com, jsaude@alumni.cmu.edu}}

\maketitle
\bibliographystyle{ecai}

\begin{abstract}
	Real data collected from different applications that have additional topological structures and connection information are amenable to be represented as a weighted graph.
    Considering the node labeling problem, Graph Neural Networks (GNNs) is a powerful tool, which can mimic experts' decision on node labeling. 
    GNNs combine node features, connection patterns, and graph structure by using a neural network to embed node information and pass it through edges in the graph. 
    We want to identify the patterns in the input data used by the GNN model to make a decision and examine if the model works as we desire. 
    However, due to the complex data representation and non-linear transformations, explaining decisions made by GNNs is challenging. 
    In this work, we propose new graph features' explanation methods to identify the informative components and important node features. 
    Besides, we propose a pipeline to identify the key factors used for node classification. 
    We use four datasets (two synthetic and two real) to validate our methods. 
    Our results demonstrate that our explanation approach can mimic data patterns used for node classification by human interpretation and disentangle different features in the graphs.
    Furthermore, our explanation methods can be used for understanding data, debugging GNN models, and examine model decisions.
\end{abstract}

\section{Introduction}

    Our contemporary society relies heavily on interpersonal/cultural relations (social networks), our economy is densely connected and structured (commercial relations, financial transfers, supply/distribution chains). 
    Moreover, those complex network structures also appear in nature, on biological systems, like the brain, vascular and nervous systems, and also on chemical systems, for instance, atoms' connections on molecules.
    Since this data is hugely structured and depends heavily on the relations within the networks, it makes sense to represent the data as a graph, where nodes represent entities and the edges the connections between them.

\begin{figure}[t] 
        \centering
        \centerline{\includegraphics[width=9cm]{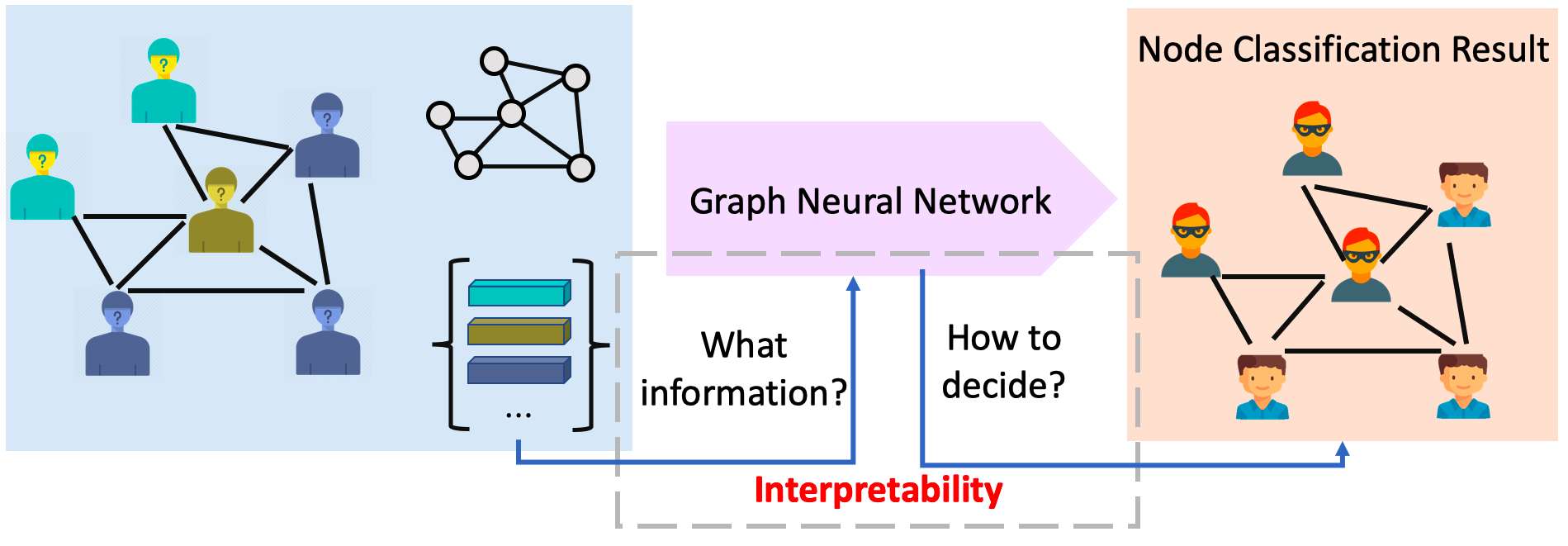}}
        \caption{Framework to explain GNN node classification.}
        \label{fig:problem}
    \end{figure}

    Graph Neural Networks (GNNs) such as GCN \cite{kipf2016semi}, GraphSage \cite{hamilton2017inductive}, can handle graph-structured data by preserving the information structure of graphs.
    Our primary focus is on the node labeling problem.
    Examples are fraud detection,  classification of social-networks' users,  role assignment on biological  structures, among others.
    GNNs can combine node features, connection patterns, and graph structure by using a neural network to embed the node information and pass it through edges in the graph. 
    However, due to the complex data representation and non-linear transformations performed on the data, explaining decisions made by GNNs is a challenging problem.
    Therefore we want to identify the patterns in the input data that were used by a given GNN model to make a decision and examine if the model works as we desire, as depicted in Figure \ref{fig:problem}.

    Although deep learning model visualization techniques have been developed in the convolution neural network (CNN), those methods are not directly applicable to explain weighted graphs with node features for the classification task. 
    A few work have been down on explaining GNN (\cite{pope2019explainability,baldassarre2019explainability,ying2019gnn,yang2019interpretable}). However, to our best knowledge, no work has been done on explaining comprehensive features (namely node feature, edge feature, and connecting patterns) in a weighted graph, especially for node classification problems. 
    Here we propose a few post-hoc graph feature explanation methods to formulate an explanation on nodes and edges.
    Our experiments on synthetic and real data demonstrate that our proposed methods and pipeline can generate similar explanations and evidence as human interpretation.  
    Furthermore,  that helps to understand whether the node features or graph typologies are the key factors used in GNN node classification of a weighted graph.

    \textbf{Our contribution} is summarized as follows:
    \begin{enumerate}
    	\item We propose the formula of \textbf{weight graph pattern} (learned by GNN) explanation as two perspectives:  \textit{Informative Components Detection and Node Feature Importance}.
    	\item We extend the current GNN explanation methods, which mainly focus on the undirected-unweighted graph to directed weighted graph. 
    	We adapt the well-know CNN visualization methods 
    	to GNN explanation.
    	\item We propose a pipeline, including novel evaluation methods, to find whether topological information or node features are the key factors in node classification. We also propose a way to discover group similarities from the disentangled results. 
    \end{enumerate}

\textbf{Paper structure:} 
In section \ref{sec:gnn}, we introduce Graph and GNN. Then in section \ref{sec:form_graph_expl}, the formula of graph explanation is described, and the corresponding methods are extended in section \ref{sec:component} and \ref{sec:feature}. In section \ref{sec:evaluation}, we propose the evaluation metrics and methods. The experiments and results are presented in section \ref{sec:experiments}. We conclude the paper in section \ref{sec:conclusion}.

\section{Graph Neural Networks}
\label{sec:gnn}
\subsection{Data Representation -- Weighted Graph} 
\label{sec:data_rep}
In this section, we introduce the necessary notation and definitions.  
We denote a graph by $G=(V,\mathcal{E})$  where $V$ 
is the set of nodes, $\mathcal{E}$ the set of edges linking the nodes and $X$ the set of nodes' features.
For every pair of connected nodes, $u, v \in V$, we denote by $e_{vu}\in \Rr$ the weight of the edge $(v,u)\in \mathcal{E}$ linking them. We denote $ E[v,u] = e_{vu}$, where $E \in \mathbb{R}^{|\mathcal{E}|}$. 
For each node, $u$, we associate a $d$-dimensional vector of features, $X_u \in \Rr^d$ and denote the set of all features as $X =\{ X_u: u \in V\} \in (\Rr^d)^{|V|}$.

Edge features contain important information about graphs. 
For instances, the graph $G$ may represent a banking system, where the nodes $V$ represents different banks, and the edges $E$ are the transaction between them; or graph $G$ may represent a social network, where the nodes $V$ represent different users, and the edges $E$ is the contacting frequencies between the users. 

We consider a node classification task, where each node $u$ is assigned a label $y_u \in I_C=\{0,\ldots, C-1\}$. The two explanation perspectives correspond to the informative s explanation on $E$ and $X$ of the weighted graph.

\subsection{GNN Utilizing Edge Weight}
Different from the state of art GNN architecture, \textit{i.e.} graph convolution networks (GCN) \cite{kipf2016semi} and graph attention networks (GAT) \cite{velivckovic2017graph}, some GNNs can exploit the edge information on graph \cite{gong2019exploiting,shang2018edge,yang2019interpretable}. 
Here, we consider weighted and directed graphs, and develop the graph neural network that uses both nodes and edges weights, where edge weights affect message aggregation. 
Not only our approach can handle directed and weighted graphs but also preserves edge information in the propagation of GNNs.
Preserving and using edges information is important in many real-world graphs such as banking payment network, recommendation systems (that use social network), and other systems that heavily rely on the topology of the connections.
Since, apart from node (atomic) features also attributes of edges (bonds) are important for predicting local and global properties of graphs.
Generally speaking, GNNs inductively learn a node  representation  by  recursively  aggregating  and  transforming the feature  vectors of its neighboring nodes. 
Following \cite{battaglia2018relational,zhang2018deep,zhou2018graph}, a per-layer update of the GNN in our setting involves these three computations, message passing Eq. \eqref{msg}, message aggregation Eq. \eqref{agg}, and updating node representation Eq. \eqref{update}, which can be expressed as:
\begin{align}
	\mathbf{m}_{vu}^{(l)}& = \text{MSG}(\mathbf{h}_u^{(l-1)},\mathbf{h}_v^{(l-1)},e_{vu}) \label{msg}\\
	\mathbf{M}_i^{(l)}   & = \text{AGG}(\{\mathbf{m}_{vu}^{(l)},e_{vu}\}\mid v \in \mathcal{N}(u)\}) \label{agg}\\
	\mathbf{h}_u^{(l)}   & = \text{UPDATE}(M_u^{(l)},\mathbf{h}_u^{(l-1)}) \label{update}
\end{align}
where $\mathbf{h}_u^{(l)}$ is the embedded representation of node $u$ on the layer $l$; 
$e_{vu}$ is the weighted edge pointing from  $v$ to $u$; $\mathcal{N}(u)$ is $u$'s neighborhood from where it collects information to update its aggregated message $M_i$. 
Specifically, $\mathbf{h}_u^{(0)} =\mathbf{x}_u$ as initial, and $\mathbf{h}_u^{(L)} $ is the final embedding for node $u$ of an $L$-layer GNN node classifier.

Here, following \cite{schlichtkrull2018modeling}, we set $h^{(l)} \in \mathbb{R}^{d^{(l)}}$ and define the propagation model for calculating the forward-pass update of node representation as: 
\begin{equation} \label{GNN}
	\begin{split}
		\mathbf{h}_u^{(l)} = \sigma & \Big( W^{(l-1)}_0\mathbf{h}_u^{(l-1)}\\ 
		& +\sum_{v\in \mathcal{N}(u)}\phi\lb W^{(l-1)}_1 \mathbf{h}^{(l-1)}_v,\mathbf{h}_u^{(l)}, e_{vu}\rb \Big) ,
	\end{split}
\end{equation}
where $\mathcal{N}(u)$ denotes the set of neighbors of node $u$ and $e_{vu}$ denotes the directed edge from $v$ to $u$, $W$ denotes the model's parameters to be learned, and $\phi$ is any linear/nonlinear function that can be applied on neighbour nodes' feature embedding.
$d^{(l)}$ is the dimension of the $l^{th}$ layer representation. 

Our method can deal with negative edges-weighted by re-normalizing them to a positive interval, for instances $[0,1]$, therefore in the following we use only positive weighted edges.
Hence, in the existing literature and in different experiment setting based on the natural of input graph, edge weights normally can play two roles: 1) message filtering; and 2) node embedding. 

\subsubsection{Type I: Edge Weights for Message Filtering} 
As the graph convolution operations in \cite{gong2019exploiting}, the edge feature matrices will be used as filters to multiply the node feature matrix. 
The GNN layer using edge weight for filtering can be formed as the following steps:
\begin{align}
	\mathbf{m}_{vu}^{(l)}& = W^{(l-1)}_1\mathbf{h}_v^{(l-1)}&\text{(message)}\label{msg2}\\
	\mathbf{M}_u^{(l)} & = \sum_{v \in \mathcal{N}(u)}\textit{g}( \mathbf{m}_{vu}^{(l)},\mathbf{h}_u^{(l-1)},e_{vu})&\text{  (aggregate)} \label{agg2}\\
	\mathbf{h}_u^{(l)} & = \sigma(W^{(l-1)}_0\mathbf{h}_u^{(l-1)}+\mathbf{M}_u^{(l)})&\text{(update)} \label{update2}
\end{align}
To avoid increasing the scale of output features by multiplication, the edge features need to be normalized, as in GAT \cite{velivckovic2017graph} and GCN \cite{kipf2016semi}. 
Due to the aggregation mechanism, we normalize the weights by in-degree $\bar{e}_{vu}= e_{vu}/\sum_{v\in\mathcal{N}(u)}e_{vu}$. 
Depending on the the problem:
\begin{itemize}
	\item \textit{g} can simply defined as: \textit{g} $= \bar{e}_{vu}\mathbf{m}_{vu}^{(l)}$; or
	\item \textit{g} can be a gate function, such as a rnn-type block of $\mathbf{m}_{vu}^{(l)}$, i.e. $\textit{g} = GRU(\bar{e}_{vu}\mathbf{m}_{vu}^{(l)},\mathbf{h}_u^{(l-1)}) $.
\end{itemize} 

\subsubsection{Type II: Edge Weights for Node Embedding} 
If $e_{vu}$ contributes to node $u$'s feature embedding, $g = f(e_{vu})\mathbf{m}_{vu}^{(l)}$, where $f(e_{vu})$ is composition of one fully-connected (FC) layer and reshape operation, mapping  $\mathbb{R} \mapsto  \mathbb{R}^{d^{(l)} \times d^{(l-1)}}$. 
In this case, we will replace equation (\ref{msg2}) and (\ref{agg2}) by:
\begin{align}
	\mathbf{m}_{vu}^{(l)}& = f(e_{vu})\mathbf{h}_v^{(l-1)}&\text{(message)}\label{msg3}\\
	\mathbf{M}_u^{(l)} & = \sum_{v \in \mathcal{N}(u)}\textit{g}( \mathbf{m}_{vu}^{(l)},\mathbf{h}_u^{(l-1)})&\text{  (aggregate)} \label{agg3}
\end{align}
Similarly, \textit{g} can be $\textit{g}=\mathbf{m}_{vu}^{(l)}$ or $g = GRU(\mathbf{m}_{vu}^{(l)},\mathbf{h}_u^{(l-1)}) $.

For the final prediction, we apply an Fully Connected (FC) layer:
\begin{equation}\label{fcn}
	\hat{\mathbf{y}}_u = \textit{softmax}(W_c\mathbf{h}_u^{(L)} + b_c)
\end{equation}

Since Type II can be converted to unweighted graph explanation, which has been studied in existing literature \cite{ying2019gnn,pope2019explainability}, the following explanation will focus on Type I. 
For generalizations, we focus on model agnostic and post-hoc explanation, without retraining GNN and modifying pre-trained GNN architectures.

\section{Formula of Graph Explanation} \label{sec:form_graph_expl}

We consider the weighted graph feature explanation problem as a two-stage pipeline.

First, we train a node classification function, in this case, a GNN.
The GNN inputs are a graph $G=(V, \mathcal{E})$, its associated node feature $X$ and its true nodes labels $Y$.  
We represent this classifier as $\Phi: G \mapsto (u \mapsto y_u)$, where $y_u \in I_C$.
The advantage of the GNN is that it keeps the flow of information across nodes and the structure of our data. 
Furthermore, it is invariant to permutations on the ordering. Hence it keeps the relational inductive biases of the input data (see \cite{battaglia2018relational}).

Second, given the node classification model and node's true label 
, the explanation part will provide a subgraph and a subset of features retrieved from the $k$-hop neighborhood of each node $u$, for $k\in \Nn$ and $u\in V$.  
Theoretically, the subgraph, along with the subset of features is the minimal set of information and information flow across neighbor nodes of $u$, that the GNN used to compute the node's label.

We define $G_S = (V_S, \mathcal{E}_S)$ to be a subgraph of $G$, where $G_S \subseteq G$, if $V_S \subseteq V$ and $\mathcal{E}_S \subseteq \mathcal{E}$. Consider the classification $y_u\in I_C$ of node $u$, then our Weighted Graph Explanation methods has two explanation components:
\begin{itemize}
	\item \textbf{Informative Components Detection}. Our method computes a subgraph, $G_S$, containing $u$, that aims to explain the classification task by looking at the edge connectivity patterns $\mathcal{E}_S$ and their connecting nodes $V_S$. 
	This provides insights on the characteristics of the graph that contribute to the node's label.
	\item \textbf{Node feature Importance}. Our method assigns to each node feature a score indicating its importance and ranking. 
\end{itemize}
\section{Informative Components Detection}
\label{sec:component}
Relational structures in graphs often contain crucial information for node classification, such as graph's topology and information flow (i.e., direction and amplitude). 
Therefore, knowing which edges contribute the most to the information flow towards or from a node is important to understand the node classification evidence. 
In this section, we discuss methods to identify the informative components on weighted graphs.
\subsection{Computational graph}
Due to the properties of the GNN, \eqref{agg}, we only need to consider the graph structure used in aggregation, \textit{i.e.} the \textit{computational graph} w.r.t node $u$ is defined as $G_c(u)$ containing $N'$ nodes, where $N'\leq N$. 
The node feature set associated with the $G_c(u)$ is $X_c(u) = \{x_v|v \in V_c(u)\}$. 
The prediction of GNN $\Phi$ is given by $\hat{y}_u=\Phi(G_c(u),X_c(u))$, which can be considered as a distribution $P_{\Phi}(Y|G_c,X_c)$ mapping by GNN.
Our goal is to identity a subgraph $G_S \subseteq G_c(u)$ (and its associated features $X_S = \{\mathbf{x}_w|w \in V_S\}$, or a subset of them) which the GNN uses to predict $u$'s label. 
In the following subsections, we introduce three approaches to detect explainable components within the computational graph: 1) Maximal Mutual Information (MMI) Mask; and 2) Guided Gradient Salience. 

\subsection{Maximal Mutual Information (MMI) Mask}    
    We first introduce some definitions.
    We define the Shannon entropy of a discrete random variable, $X$, by $H(Y)= \mathbb{E} [-\log(P(X))]$, where $P(X)$ is the probability mass function. 
    Furthermore, the conditional entropy is defined as:
    \[
        H[Y|X] = - \sum_{x\in \mathcal{X}, y \in \mathcal{Y}} p(x,y) \log \frac{p(x,y)}{p(x)},
    \]
    where $\mathcal{X}$ and $\mathcal{Y}$ are the sample spaces. 
    Finally, we define the mutual information (MI) between two random variables as $I(Y,X)~=~H(Y)- H(Y|X)$, this measures the mutual dependence between both variables. 

    Using ideas from Information theory \cite{cover2012elements} and following GNNExplainer \cite{ying2019gnn}, the informative explainable subgraph and nodes features subset are chosen to maximize the mutual information (MI):
\begin{equation} \label{MI1}
	\max\limits_{G_S} I(Y,(G_S,X_S))\!=\!H(Y|G,X)\!-\!H(Y|G_S,X_S)
\end{equation}
Since the trained GNN node classifier $\Phi$ is fixed, the $H(Y)$ term of Eq.(\ref{MI1}) is constant. 
As a result, it is equivalent to minimize the conditional entropy $H(Y|G_S,X_S)$.
\begin{equation} \label{MI2}
	\begin{split}
		-\mathbb{E}_{Y|G_S,X_S}[\log P_{\Phi}(Y|G_S,X_S)]   
	\end{split}
\end{equation}
Therefore, the explanation to the graph components with prediction power w.r.t node $u$'s prediction $\hat{y}_u$ is a subgraph $G_S$ and its associated feature set $X_S$, that minimize \eqref{MI2}. 
The objective of the explanation thus aims to pick the top informative edges and its connecting neighbours, which form a subgraph, for predicting $u$'s label.
Because, probably some edges in $u$'s computational graph $G_c(u)$ form important message-passing \eqref{agg2} pathways, which allow useful node information to be propagated across $G_c(u)$ and aggregated at $u$ for prediction; while some edges in $G_c(u)$ might not be informative for prediction.
Instead of directly optimize $G_S$ in Eq. (\ref{MI2}), as it is not tractable and there are exponentially many discrete structures $G_S \subseteq G_c(u)$ containing $N^\prime$ nodes, GNNExplainer \cite{ying2019gnn} optimizes a mask $\mathcal{M}_{sym}^{N^\prime \times N^\prime}[0,1]$ on the binary adjacent matrix, which allows gradient descent to be performed on $G_S$.

If the edge weights are used for node embedding, the connection can be treated as binary and fit into the original GNNExplainer.However, if edge weights are used as filtering, the mask should affect filtering and normalization. We extend the original GNNExplainer method by considering edge weights and improving the method by adding extra regularization. 
Unlike GNNExplainer, where there are no constraints on the mask value, we add constraints to the value learned by the mask
\begin{equation}
	\begin{cases}
		\sum_w \mathcal{M}_{vw}e_{vw} = 1 \\
		\mathcal{M}_{vw} \geq 0, & \text{ for $(v,w) \in \mathcal{E}_c(u)$}
	\end{cases}
\end{equation}
and perform a projected gradient decent optimization. 
Therefore, rather than optimizing a relaxed adjacency matrix in GNNExplainer, we optimize a mask $\mathcal{M} \in [0,1]^Q $ on weighted edges, supposing there are Q edges in $G_c(u)$. 
Then $E^{\mathcal{M}}_c = E_c \odot \mathcal{M}$, where $\odot$ is element-wise multiplication of two matrix. 
The masked edge $E^{\mathcal{M}}_c$ is subject to the constraint that $E^{\mathcal{M}}_c[v,w]\leq E_c[v,w]$, $\forall(v,w) \in \mathcal{E}_c(u)$.
Then the objective function can be written as:
\begin{align} \label{obj3}
	\min\limits_{M}-\sum_{c=1}^C\mathbb{I}[y=c]\log P_{\Phi}(Y|G_c = (V_c, E_c \odot \mathcal{M}), X_c)  
\end{align}

In GNNExplainer, the top $k$ edges may not form a connected component including the node (saying $u$) under prediction i. 
Hence, we added the entropy of the $(E_c\odot \mathcal{M})_{vu}$ for all the node $v$ pointing to node $u$' as a regularization term, to ensure that at least one edge connected to node $u$ will be selected. 
After mask $\mathcal{M}$ is learned, we use threshold to remove small $E_c\odot \mathcal{M}$ and isolated nodes. 
Our proposed optimization methods to optimize $\mathcal{M}$ maximizing mutual information (equation \eqref{MI1}) under above constrains is shown in Algorithm \ref{ag1}.
\begin{algorithm}[t]
	\caption{ Optimize mask for weighted graph}\label{ag1}
	\hspace*{\algorithmicindent} \textbf{Input:} {1. $G_c(u)$, computation graph of node $u$;  2. Pre-trained GNN model $\Phi$;  3. $y_u$, node $u$'s real label; 4. $\mathcal{M}$, learn-able mask; 5. $K$, number of optimization iterations; 6. $L$, number of layers of GNN.
		\begin{algorithmic}[1]
			\State {$\mathcal{M} \gets \text{randomize parameters} $}
			\Comment{initialize, $\mathcal{M} \in [0,1]^Q$}
			\State{$\mathbf{h}_v^{(0)} \gets \mathbf{x}_v$, for $v\in G_c(u)$}
			\For{$k = 1$ to $K$}
			\State{$\mathcal{M}_{vw} \gets \frac{\textit{exp}(\mathcal{M}_{vw}e_{vw})}{\sum_{v}\textit{exp}(\mathcal{M}_{vw}e_{vw})}$}
			\Comment{renormalize mask}				
			\For{$l = 1$ to $L$}
			\State{$\mathbf{m}_{vu}^{(l)} \gets W_1^{(l-1)} \mathbf{h}_{v}^{(l-1)}$}
			\Comment{message}
			\State{$M_u^{(l)} \gets \sum_v  \textit{g}(\mathcal{M}_{vu}\mathbf{m}_{vu}^{(l)},\mathbf{h}_{u}^{(l-1)}) $}
			\Comment{aggregate}
			\State{$\mathbf{h}_u^{(l)} \gets \sigma(W_0\mathbf{h}_u^{(l-1)}+M_u^{(l)}) $}
			\Comment{update}
			\EndFor
			\State{$\hat{\by}_u \gets \textit{softmax}(\mathbf{h}_u^{(L)})$}
			\Comment{predict on masked graph}
			\State {$loss \gets \textit{crossentropy}(\by_u, \hat{\by}_u)+ \textit{regularizations}$ } 
			\State{$\mathcal{M} \gets \textit{optimizer}(loss,\mathcal{M})$}  
			\Comment{update mask}
			\EndFor	
		\end{algorithmic}
		\hspace*{\algorithmicindent} \textbf{Return:} $\mathcal{M}$
	}
\end{algorithm}

\subsection{Guided Gradient (GGD) Salience}
Guided gradient-based explanation methods \cite{simonyan2013deep} is perhaps the most straight forward and easiest approach. 
By calculating the differentiate of the output w.r.t the model input then applying norm, a score can be obtained. 
The gradient-based score can be used to indicate the relative importance of the input feature since it represents the change in input space which corresponds to the maximizing positive rate of change in the model output. Since edge weights are variables in GNN, we can obtain the edge mask as 
\begin{equation}
	g_{vu}^E = \textit{ReLU} \left(\frac{\partial \hat{y}_u^c}{\partial e_{vu}} \right)
\end{equation}
where $c \in \{0,\ldots, C-1\}$ is the correct class of node $u$, and $y^u_c$ is the score for class $c$ before softmax layer. 
where $\mathbf{x}_{v}$ is node $v$'s feature. We normalize $ g_{vu}^E$ by dividing  $ max(g_{vu}^E)$ to be bound it to $[0,1]$.
Here, we select the edges whose $g^E$ is in the top $k$ largest ones and their connecting nodes. 
The advantage of contrasting gradient salience method is easy to compute. 

\section{Node Feature Importance}
\label{sec:feature}
Node's features information play an important role in computing messages between nodes.
That data contribute to the message passing among nodes in the message layer (see Eq. \eqref{msg}).
Therefore, the explanation for the classification task (or others, like regression) must take into account the feature information. 
In this section, we will discuss three approaches to define node feature importance in the case that the node attribute $X_u\in \Rr^d$ is a vector containing multiple features.

\subsection{Maximal Mutual Information (MMI) Mask}    
Following GNNExplainer \cite{ying2019gnn}, in addition to learning a mask on edge to maximize mutual information, we also can learn a mask on node attribute to filter features given $G_S$. The filtered node feature $X_S^T = X_S \odot \mathcal{M}_T$, where $\mathcal{M}_T$ is a feature selection mask matrix to be learned, is optimized by
\begin{equation*}
	\min\limits_{M_T}-\sum_{c=1}^C\mathbb{I}[y=c]\log P_{\Phi}(Y|G_S, X_S \odot\mathcal{M}_T))   
\end{equation*}
In order to calculate the output given $G_S$ but without feature $T$ and also guarantee propagation, a reparametrization on $X$ is used in paper \cite{ying2019gnn}:
\begin{equation}
	X = Z + (X_S -Z) \odot \mathcal{M}_T, \quad s.t.\sum_j\mathcal{M}_{Tj}<k    
\end{equation}
where $Z$ is a matrix with the same dimension of $X_S$ and each column $i$ is sampled from the Gaussian distribution with mean and std of the $i_{th}$ row of $X_S$. To minimize the objective function, when $i_{th}$ dimension is not important; that is, any sample of $Z$  will pull the corresponding mask value $\mathcal{M}_{Ti}$ towards 0; if $i_{th}$ dimension is very important, the mask value $\mathcal{M}_{Ti}$ will go towards 1. Again, we set constrain:
\begin{equation}
	0\leq \mathcal{M}_{Ti} \leq 1, 
\end{equation}
and perform projected gradient decent optimization.

However, before performing optimization on $\mathcal{M}_T$, $Z$ is only sampled once. 
Different samples of $Z$ may affect the optimized $\mathcal{M}_T$, resulting in unstable results. Performing multiple sampling of $Z$ will be time-consuming since each sample is followed by optimization operation on $\mathcal{M}_T$.

\subsection{Prediction Difference Analysis (PDA)}
We propose using PDA for node features importance, which can cheaply perform multiple random sampling with GNN testing time. 
The importance of a nodal feature, towards the correct prediction, can be measured as the drop of prediction score to its actual class after dropping a certain nodal feature. 
We denote by $X{\setminus i}$ the subset of the feature set $X$ where we removed feature $x_i$.
The prediction score of the corrupted node is $P_{\Phi}(y=y_u | G = G_S, X = X_S{\setminus i})$. 
To compute $P_{\Phi}(y=y_u | G = G_S, X = X_S{\setminus i})$, we need to marginalize out the feature $x_i$: 
\begin{equation}
	\bar{P}=\mathbb{E}_{{\hat{x}_i}\sim p({x_i}|{X_S{\setminus i}})}P_{\Phi}(y=y_u|G = G_S, X = \{X_S{\setminus i},\hat{x}_i\}),
\end{equation}
Modeling $p({x_i}|{X_S{\setminus i}})$ by a generative model can be computationally intensive and may not be feasible.  We empirically sample $\hat{x}_i$ from training data. Noting that the training data maybe unbalance, to reduce sampling bias we should have $p({x_i}\in K|{X_S{\setminus i}}) \propto 1/N_{k}$, where $K$ is the features space of class $k$ and $N_k$ is the number of training instance in class $k$. Explicitly, $p({x_i}\in K|{X_S{\setminus i}}) \propto 1/N_{k}$. We define the importance score for $i_{th}$ node feature as the difference of original prediction score
\begin{equation}
	PDA_i  = \textit{ReLU}(P_{\Phi}(y=y_u | G = G_S, X = X_S)  - \bar{P}).
\end{equation}
Naturally, $PDA_i$ is bounded in $[0,1]$. The larger the $PDA_i$ indicates a more important the $i_{th}$ feature.
\subsection{Guided Gradient (GGD) Node Feature Salience}
Similar to the guided gradient method in detecting explainable components, we calculate the differentiate of the output with respect to the node under prediction and its neighbors in its computation graph $G_c(u)$ on the $i_{th}$ feature for $i\in I_C$: 
\begin{align}
	g_v^i &= \textit{ReLU}\lb \frac{\partial \hat{y}_u^c}{\partial x_v^i} \rb, \quad v \in G_c(u). 
\end{align}
The larger the $g_i$ is, the more important the $i_{th}$ feature is.

\section{Evaluation Metrics and Methods}
\label{sec:evaluation}
For synthetic data, we can compare explanation with data generation rules. However, for real data, we do not have ground truth for the explanation. In order to evaluate the results, we propose the evaluation metrics for quantitatively measuring the explanation results and propose the correlation methods to validate if edge connection patter or node feature is the crucial factor for classification. 
\begin{figure}[t] 
    \centering
    \centerline{\includegraphics[width=9cm]{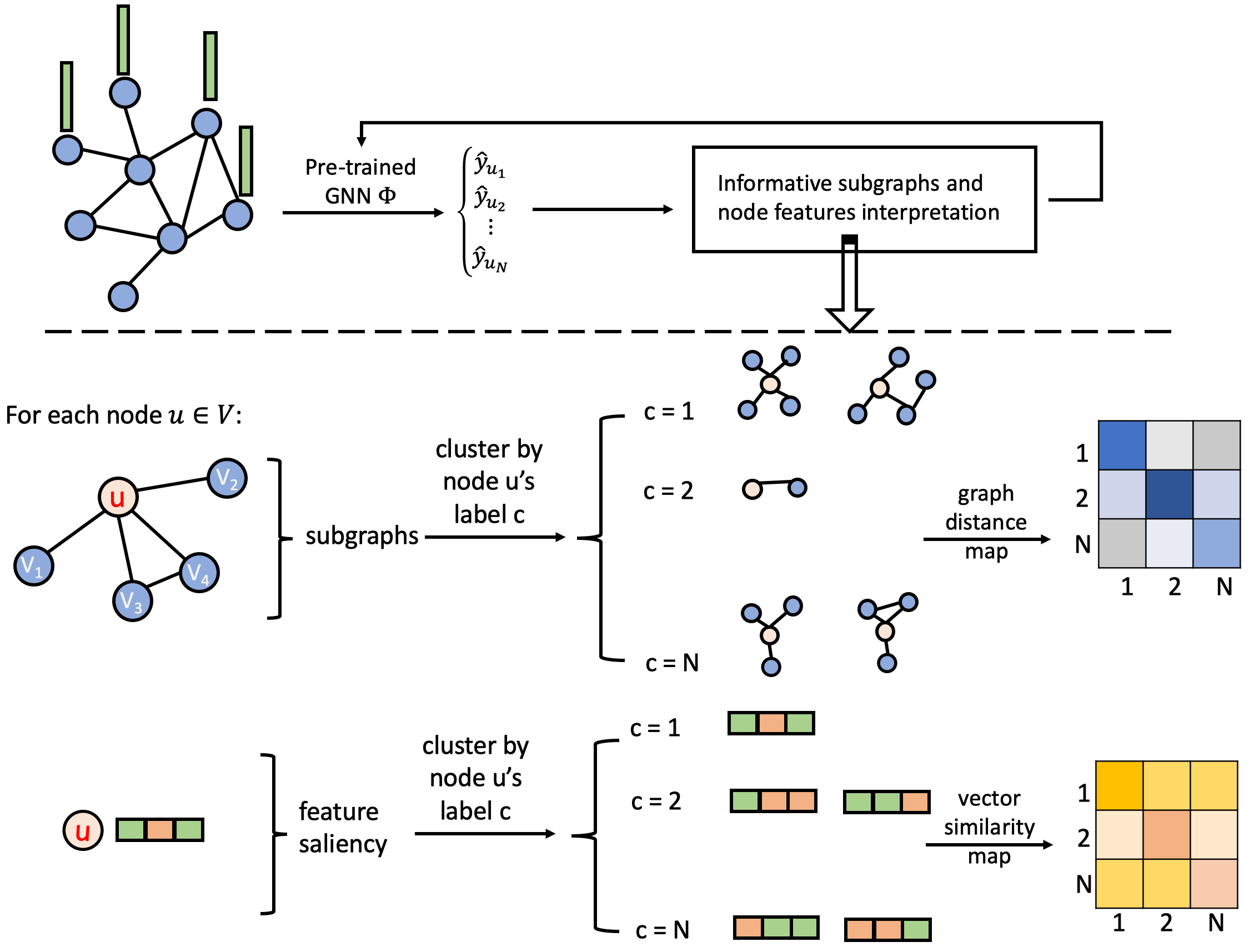}}
    \caption{Disentangle informative subgraphs and node features,}
    \label{fig:correlation}
\end{figure}

\subsection{Evaluation Metrics}
We define metrics \textit{consistency}, \textit{contrastivity} and \text{sparsity} (Here, definition of \textit{contrastivity} and\text{sparsity} are different from the ones in\cite{pope2019explainability}) to measure informative component detection results. Firstly, To measure the similarity between graphs, we introduce graph edit distance (GED) \cite{abu2015exact}, which is a graph similarity measure analogous to Levenshtein distance for strings. It is defined as minimum cost of edit path (sequence of node and edge edit operations) transforming graph G1 to graph isomorphic to G2. In case the structure is isomorphic but edge weights are different. If GED=0, Jensen-Shannon Divergence (JSD) \cite{nielsen2010family},  is added on GED to further compare the two isomorphic subgraphs. 
Specifically, we design consistency as the GED between the informative subgraphs of the node in the same class,     as whether the informative components detected for the node in the same class are consist; and design contrastivity as the GED across the informative subgraphs of the node in the same class, as and whether the informative components detected for the node in the different class are contrastive; Sparsity is defined as the density of mask $\sum_{e_{vw}\in G_c(u)} \Upsilon_{vw}/Q, \Upsilon \in \{\mathcal{M}, g^E\}$, as the density of component edge importance weights. 

\subsection{Important features disentanglement}
    We follow the pipeline described in Figure \ref{fig:correlation}.
    Hence, after training a GNN, we perform informative component detection and node importance analysis on each of the nodes $u\in V$. 
    Furthermore, we get the local topology $G_S(u)$ that explains the labeling of that node.
    After, for each label $c\in I_C$, we collect all the subgraphs that explain that label, $\{G_S(w)\}_{w\in c}$, where $c\in I_C$ means that node $w$ is classified as class $c$.
    Then, we measure the distance, using the predefined GED, from all the subgraphs in each label $c$ to all the subgraphs in all labels $j \in I_C$.
    So,  we obtain a set of distances between the instance within the class and across classes. 
    Similarly, for each label $c\in I_C$, we collect all the node feature  saliency vectors that explain that label, $\{\mathcal{F}(w)\}_{w\in c}$, where $c\in I_C$ means that node $w$ is classified as class $c$ ,and $\mathcal{F} \in \{\mathcal{M}_T, PDA, g\}$.
    We then measure the similarity using predefined Pearson correlation of all the feature saliency vectors in each label $c \in I_C$, so that we obtain a set of correlations between the instance within the class and across classes.
        
    As the last step, we group the distance and correlations by class-pairs and take the average of the instance in each class pair. 
    Therefore, we generate a $C\times C$ distance map for informative components and a $C\times C$ similarity map for node feature salience. 
    The key features should have high consistency within the groups and contrastivity across different classes. 
    Therefore, we examine the distance map and similarity map of the given graph and GNN classifier. 
    If topology information contributes significantly to the GNN, the diagonal entries of distance maps should be small, while the other entries should be large. 
    When node features are key factors for node labeling, the diagonal entries of distance maps should be large, while the other entries should be small. 
    From those maps, not only we can examine if the detected informative components or the node features are meaningful for node classification, but also we find which classes have similar informative components or important node features.

\section{Experiments}
\label{sec:experiments}
\begin{figure}[t] 
	\begin{minipage}[b]{0.9\linewidth}
		\centering
		\centerline{\includegraphics[width=8cm]{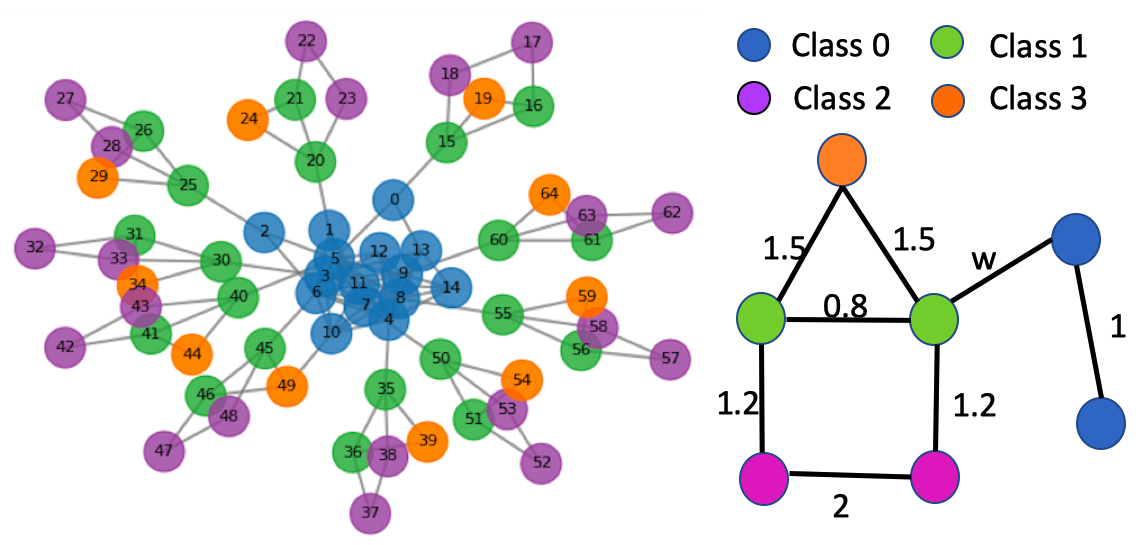}}
	\end{minipage}
	\caption{Synthetic BA-house graph data and corresponding edge weights, each BA node belongs to class "0," and each "house" shape node belongs labeled "1-3" based on its motif. The node orders are denoted. }
	\label{fig:syn}
\end{figure} 

    The topic that we addressed in this work of model-agnostic GNN post-hoc  explaination was quite new.  
Few previous studies could be compared to our work.  For example, Pope et al.  \cite{pope2019explainability} formulated GNN differently, which replied on adjacent matrix, and the attention method in \cite{ying2019gnn} is model-specific. Therefore,  those methods were not easily adopted.  We mainly compared with the original MMI Mask proposed in GNNExplainer \cite{ying2019gnn}. Furthermore, to our knowledge, graph feature importance disentangle pipeline is first proposed here.
    We simulated synthetic data and compared the results with human interpretation to demonstrate the feasibility of our methods. 
    Note that, the color codes for all the figures below follow the on denoted in Figure \ref{fig:syn}. 
    The red node is the node we try to classify and explain.
\subsection{Synthetic Data 1 - SynComp}

Following \cite{ying2019gnn}, we generated a Barabási–Albert (BA) graph with $15$ nodes and attached $10$ five-node house-structure graph motifs are attached to random nodes, ended with 65 nodes in Figure \ref{fig:syn}. We created a small graph for visualization purpose. However, the experiment results held for large graphs. Several natural and human-made systems, including the Internet, citation networks, social networks, and banking payment system can be thought to be approximately a BA graph, which certainly contains few nodes (hubs) with unusually high degree and a big number of nodes poorly connected. The edges connecting with different node pairs were assigned different weights denoted in Figure \ref{fig:syn} as well, where $w$ was an edge weight we will discuss later. Then, we added noise to synthetic data by uniformly randomly adding $0.1N$ edges, where $N$ was the number of nodes in the graph. 
In order to constrain the node label is determined by motif only, all the node feature $\mathbf{x}_i$ was designed the 2-D node attributes with the same constant. 

\begin{figure}[t] 
	\begin{minipage}[b]{1.0\linewidth}
		\centering
		\centerline{\includegraphics[width=8cm]{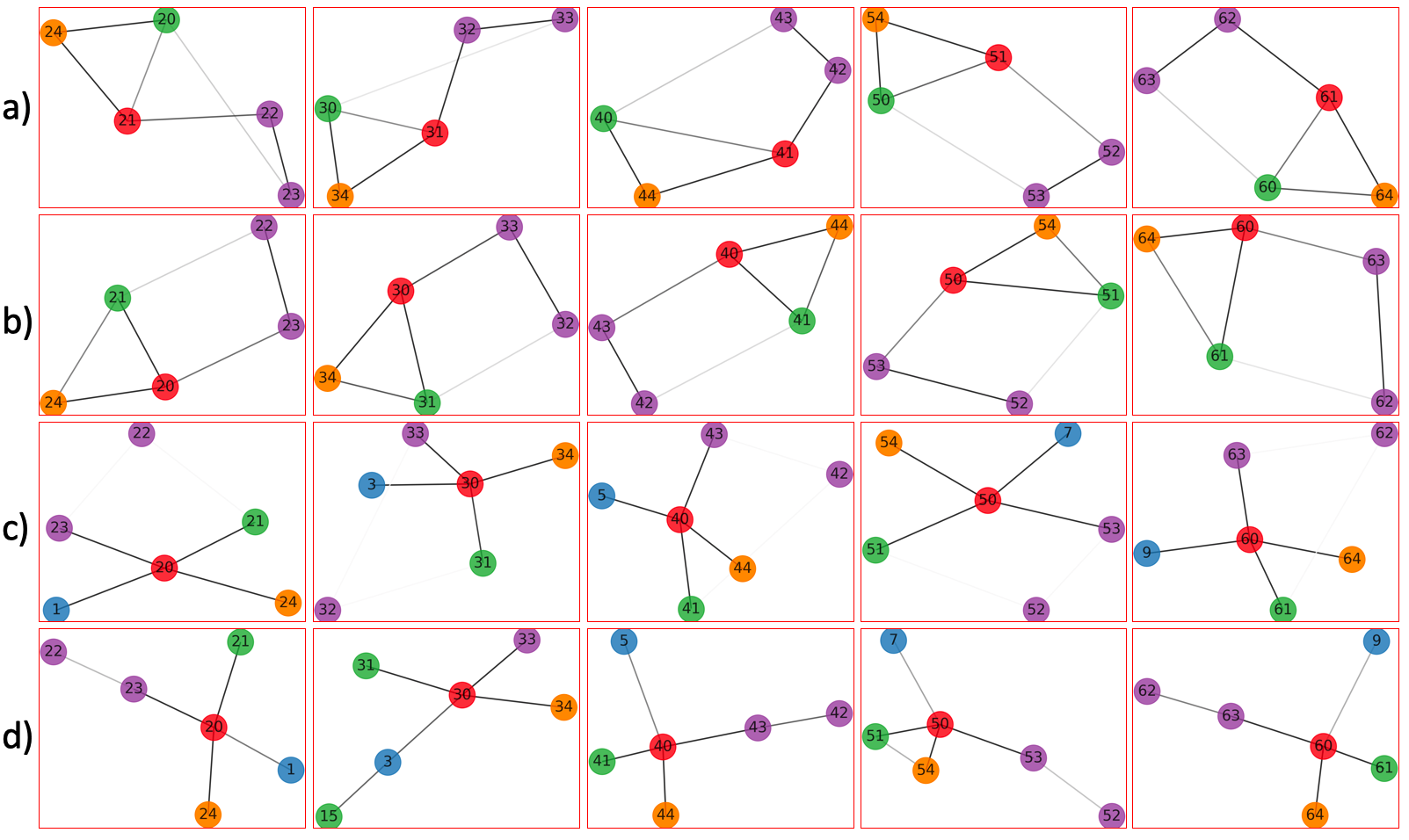}}
	\end{minipage}
	\caption{Informative Components. Row a)-c), $w=0.1$. Row a) is for the node in class one not connecting to class 0 nodes using MMI mask. Row b) is for the node in class one connecting to class 0 nodes using MMI mask. Row c) is for the node in class one connecting to class 0 nodes using GGD. 
		Row d) is for the node in class one connecting to class 0 nodes using MMI mask, but $w=2$.}
	\label{fig:syn_res}
\end{figure}

We use $\textit{g} = \bar{e}_{vu}\mathbf{m}_{vu}^{(l)}$ in Eq. \eqref{msg2}. The parameters setting are \textit{input\_dim = 2, hidden\_dim = 8, num\_layers = 3 and epoch =300.} We randomly split $60\%$ of the nodes for training and the rest for testing.
GNN achieved $100\%$ and $96.7\%$ accuracy on training and testing dataset correspondingly. We performed informative component detection (kept top 6 edges) and compare them with human interpretation -- the 'house shape,' which can be used as a reality check (Table \ref{tb:tb1}). The GNNExplainer \cite{ying2019gnn} performed worse in this case, because it ignored the weights on the edge so that the blue nodes were usually included into the informative subgraphs. In Figure \ref{fig:syn_res}, we showed the explanation results of the node in the same place but has different topology structure (row a \& b) and compared how eight weights affected the results (row a \& d). We also showed the results generated by different methods (row a \& c).

\begin{table}
	\small
	\vspace{-1em}
	\centering
	\begin{tabular}{@{}l|c c c c@{}}
		\toprule
		\textbf{Method}& MMI mask& GGD  & GNNExplainer \cite{ying2019gnn} \\ \midrule
		\textbf{AUC}& $\mathbf{0.932}$&0.899 &0.804 \\
		\bottomrule
	\end{tabular}
		\\
	(\emph{Measuring on all the nodes in class 1 with $w=0.1$})
	\caption{Saliency component compared with 'house' shape. \label{tb:tb1}}
\end{table}

\subsection{Synthetic Data 2 - SynNode} 
In order to constrain the node labels were determined by node features only, we constructed a graph with BA topology and designed the 2D node attributes for each node on the graph. We generated a random normal distributed noise $s_u$ for each node $u$, where $s_u \sim N(0,0.1)$. The $1_{st}$ entry of the node attribute vector was assigned as $s_u$. For the $2_{nd}$ entry, the value is $s_u + (y_u+1)*0.2$, where $y_u$ is the real label of $u$. We constructed a graph containing 60 nodes and randomly removed half of the edges to make it sparse. We used the same training model and methods in SynComp.
For the quantitative measurement on node importance, we calculated the accuracy of classifying the $2_{nd}$ entry as the important features. Then we applied softmax function on the node feature importance vectors and calculated their mean square error (MSE) with $[0,1]^{\intercal}$. Last, we theoretically listed the computation complexity estimation. We show the measurements on one example node in Table \ref{tb:tb2}, where $k$ is number of sampling times.
\begin{table}
	\small
	\centering
	\begin{tabular}{@{}p{1.5cm}|c c c@{}}
		\toprule
		\textbf{Method}& MMI mask& PDA &GGD\\ \midrule
		\textbf{Accuracy} & $100 \pm 0 \% $ & $100 \pm 0\%$ &$100 \pm 0\%$\\
		\textbf{MSE} &  $0.29 \pm 0.03$& $0.30 \pm 4e^{-4}$  &  $0.17\pm 0.00$ \\
		\textbf{Time cost} & Train  & $k$Test & Test \\\toprule
	\end{tabular}
	(\emph{Repeating 10 times, mean $\pm$ std})
	\caption{Compare importance score with ground truth. \label{tb:tb2}}
\end{table}



\subsection{Citation Network Data} 
\begin{figure}[t]
	\begin{minipage}[b]{1.0\linewidth}
		\centering
		\centerline{\includegraphics[width=8cm]{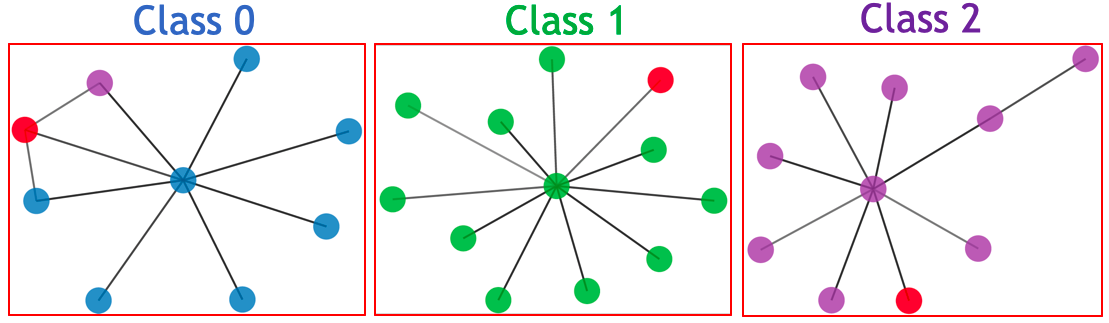}}
	\end{minipage}
	\caption{Overlapping Informative components detected by MMI mask and GGD for the examples of each class. \label{fig:pubmed_res}}
\end{figure}

PubMed dataset \cite{pubmed} contains 19717 scientific publications pertaining to diabetes classified into one of three classes, "Diabetes Mellitus, Experimental," "Diabetes Mellitus Type 1", "Diabetes Mellitus Type 2". The citation network built on PubMed consists of 44338 links. Each publication in the dataset is described by a TF/IDF weighted word vector from a dictionary which consists of 500 unique words. Edge attribute is defined as a positive Pearson correlation of the node attributes. We randomly split $80\%$ of the nodes as training data and rest as testing dataset. GNN used edge as filtering and $\textit{g} = \bar{e}_{vu}\mathbf{m}_{vu}^{(l)}$. The parameters setting are \textit{ hidden\_dim = 32, num\_layers = 3 and epoch =1000}. Learning rate was initialized as 0.1, and decreased half per 100 epochs. We achieved an accuracy of 0.786 and 0.742 on training and testing data separately. We selected top 20 edges in both MMI and GGD, show the overlapping informative component detection results of an example in each class in Figure \ref{fig:pubmed_res}. Obviously, we can find the pattern that those nodes were correctly classified since they connect to the nodes in the same class.

For the selected examples, we used above three node feature importance methods to vote the top 10 important features. Specifically, we first ranked the feature (keywords in the publications) importance by each method. Different nodes' feature might have different ranks by different methods. Then we summed the rank of each feature over the three methods. The smaller the summed rank number is, the more important the feature is. The top 10 ranked keywords are  \textit{"children", "type 2", "iddm", "type 1", "insulindepend", "noninsulindepend", "autoimmun", "hypoglycemia", "oral", "fast"}. We consulted 2 diabetes experts and got the validation that \textit{"type 2", "iddm", "noninsulindepend"} were directly related to publications of class '"Diabetes Mellitus Type 2"; \textit{"autoimmune", "children", "hypoglycemia", "insulindepend", "type 1"} are closely associated to class "Diabetes Mellitus Type 1"; and \textit{"oral", "fast"} are the common experment methods in class "Diabetes Mellitus, Experimental". 

\subsection{Bitcoin OTC Data}
\begin{figure}[t]
	\begin{minipage}[b]{1.0\linewidth}
		\centering
		\centerline{\includegraphics[width=9cm]{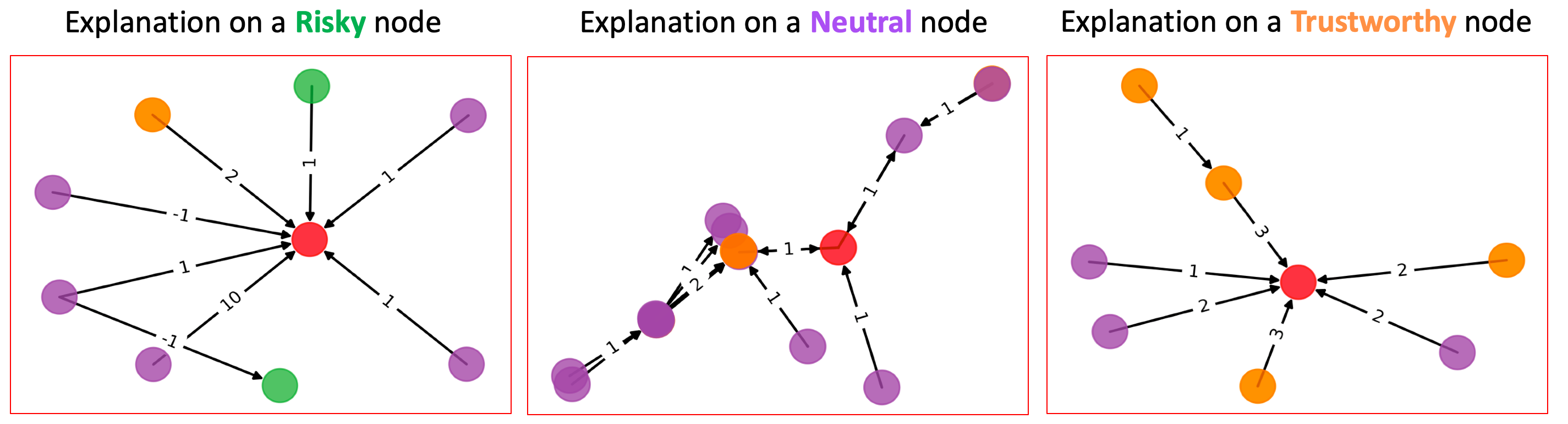}}
	\end{minipage}
	\caption{Informative subgraph detected by MMI mask (showing the original rating scores on the edges).}
	\label{fig:bitcoin_res}
\end{figure}

Bitcoin is a cryptocurrency that is used for trading anonymously. 
There is counterparty risk due to anonymity. We use Bitcoin dataset (\cite{kumar2018rev2}) collecting in one month, where Bitcoin users rate the level of trust to the users they made transactions to. 
The rating scales are from -10 to +10 (except for 0). According to OTC's guideline, the higher the rating, the more trustworthy. 
We labeled the users whose rating score had at list one negative score as risky; the users whose more than half received ratings were greater than one as trustworthy users; the users who did not receive any rating scores as an unknown group; and the rest of the users were assigned to the neural group. We chose the rating network data at a time point, which contained 1447 users, 5739 rating records.
We renormalized the edge weights to $[0,1]$ by $\tilde e_{ij} = e_{ij}/20+1/2$. Then we trained a GNN on $90\%$ unknown, neutral and trustworthy node, $20\%$ risky node, those nodes only, and perform classification on the rest of the nodes. We chose \textit{g} as a $GRU$ gate and the other settings are setting are \textit{ hidden\_dim = 32, num\_layers = 3 and epoch =1000}. Learning rate was initialized as 0.1, and decreased half per 100 epochs. We achieved accuracy 0.730 on the training dataset and 0.632 on the testing dataset. 
Finally, we showed the explanation result using MMI mask since it is more interpretable (see Figure \ref{fig:bitcoin_res}) and compared them with possible human reasoning ones. 
The pattern of the informative component of the risky node contains negative rating; 
the major ratings to a trustworthy node are greater than 1; and for the neutral node, it received lots of rating score 1. 
The informative components match the rules of how we label the nodes.

    Using both real datasets, we measured consistency, contrastivity, and sparsity by selecting the top 4 important edges. The results on the two real datasets are listed in Table \ref{tb3}.
\begin{table}
	\small
	\centering
	\begin{tabular}{@{}p{0.7cm}|p{1.2cm}|c c c@{}}
		\toprule
		& Dataset& Consistency& Contrastivity & Sparsity\\
		\midrule
		&PubMed& 2.00  & 1.99& 0.022  \\ 
		\multirow{1}{*}{\textbf{MMI}}&BitCoin& 1.81 &  2.45& 0.132 \\
		\midrule
		&PubMed& 2.14  & 2.07 & 0.049\\ 
		\multirow{1}{*}{\textbf{GGD}}&BitCoin& 2.05 & 2.60 & 0.151 \\ \bottomrule
	\end{tabular}
	(\emph{Average on 50 random correctly classified nodes in each class})
	\caption{Evaluate informative components. \label{tb3}}
\end{table}

\subsection{Feature Importance Disentanglement}
\begin{figure}[t]
        \centering
        \begin{subfigure}[b]{0.33\linewidth}
        \centering \includegraphics[width=3.3cm]{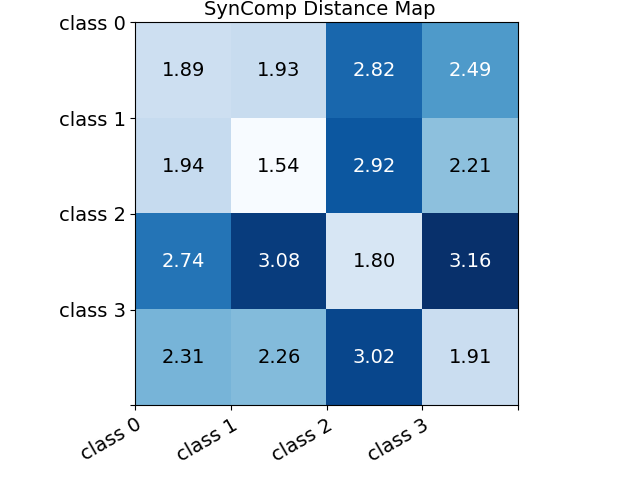}
        \caption{}
        \label{fig:dist1}
        \end{subfigure}%
        \begin{subfigure}[b]{0.33\linewidth}
        \centering \includegraphics[width=3.3cm]{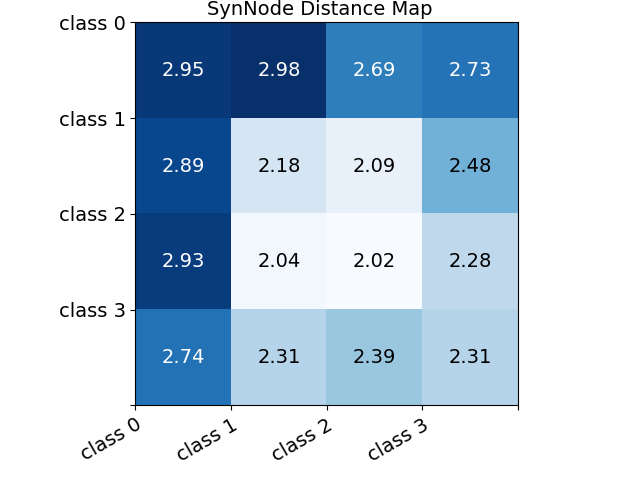}
        \caption{}
        \label{fig:dist2}
        \end{subfigure}%
            \begin{subfigure}[b]{0.33\linewidth}
        \centering \includegraphics[width=3.3cm]{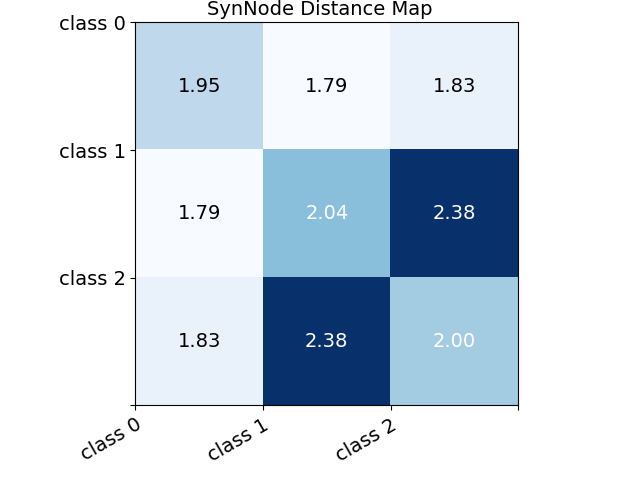}
        \caption{}
        \label{fig:dist3}
        \end{subfigure}%
        
        \begin{subfigure}[b]{0.33\linewidth}
        \centering \includegraphics[width=3.3cm]{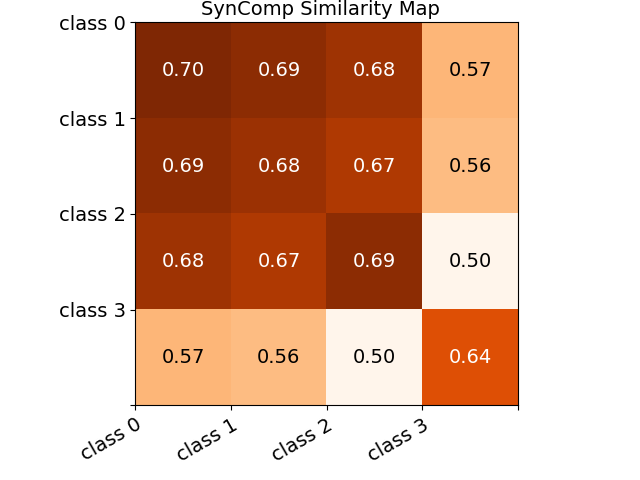}
        \caption{}
        \label{fig:sim1}
        \end{subfigure}%
        \begin{subfigure}[b]{0.33\linewidth}
        \centering \includegraphics[width=3.3cm]{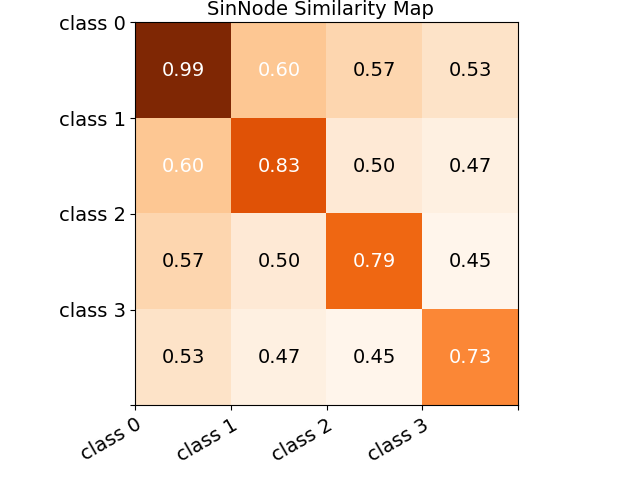}
        \caption{}
        \label{fig:sim2}
        \end{subfigure}%
            \begin{subfigure}[b]{0.33\linewidth}
        \centering \includegraphics[width=3.3cm]{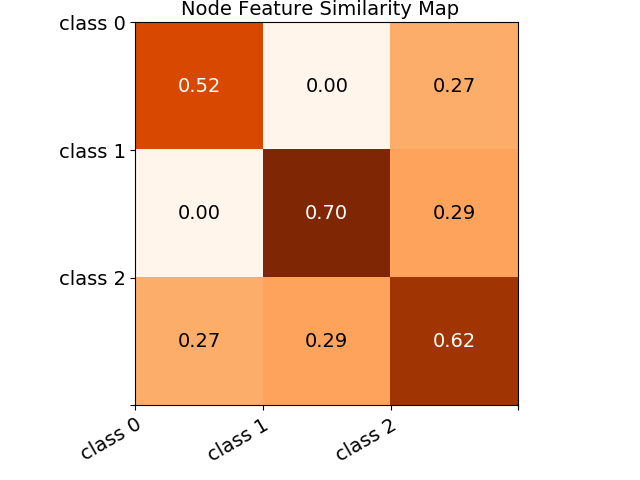}
        \caption{}
        \label{fig:sim3}
        \end{subfigure}%
        \caption{Explanation disentangle using maps: (a) SynComp Informative Subgraphs Distance Map; (b) SynNode Informative Subgraphs Distance Map; (c) PubMed Informative Subgraphs Distance Map; (d) SynComp Node Salience Similarity Map; (e) SynNode Node Salience Similarity Map; (f) PubMed Node Salience Similarity Map. }
        \label{fig:distance}
    \end{figure}

    We performed the disentanglement experiment on SynComp ($w=0.1$), SynNode and Pubmed datasets, because these datasets have both node and edge features. 
    For the Pubmed dataset, we randomly selected 50 correctly classified nodes in each class to calculate the stats. Since we had different explanation methods, we calculated the distance maps and similarity maps for each method and performed averaging over different methods.
    The distance map calculating on the subgraph with top 4 informative edges is shown in Figure \ref{fig:dist1} and \ref{fig:dist2}. 
    From the distance map, we can examine the connecting pattern is a key factor for classifying the nodes in SynComp, but  not in SynNode. 
    For the SynComp dataset, in-class distances were smaller than cross-class distances. 
    Whereas, the distance map for SynNode and PubMed did not contain the pattern. 
    Also, from the distance map, we could see the node in class 2 and 3 had the most distinguishable informative component, but classes 0 and 1's are similar. 
    For the similarity maps (Fig. \ref{fig:sim1}, Fig. \ref{fig:sim2}, and \ref{fig:sim3}), SynNode and PubMed datasets had much more significant similarities within the class compared with the similarities across the classes. 
    Combining distance maps and similarity maps for each dataset, we could understand that topology was the critical factor for SynComp dataset, and node feature was the key factor for SynNode and PubMed dataset for node classification in GNNs.
\section{Conclusion} \label{sec:conclusion}
    In this work, we formulate the explanation on weighted graph features used in GNN for node classification task as two perspectives: \textit{Components Detection} and \textit{Node Feature Importance}, that can provide subjective and comprehensive explanations of feature patterns used in GNN. 
    We also propose evaluation metrics to validate the explanation results and a pipeline to find whether topology information or node features contribute more to the node classification task.
    The explanations may help debugging, feature engineering, informing human decision-making, building trust, increase transparency of using graph neural networks, among others.
    Our future work will include extending the explanation to graphs with multi-dimensional edge features and explaining different graph learning tasks, such as link prediction and graph classification.


\bibliography{ref}

\newpage
\section*{Appendix}
  \begin{figure}[htpb]
         \begin{minipage}[b]{1\linewidth}
         		\centering
         		\centerline{\includegraphics[width=8cm]{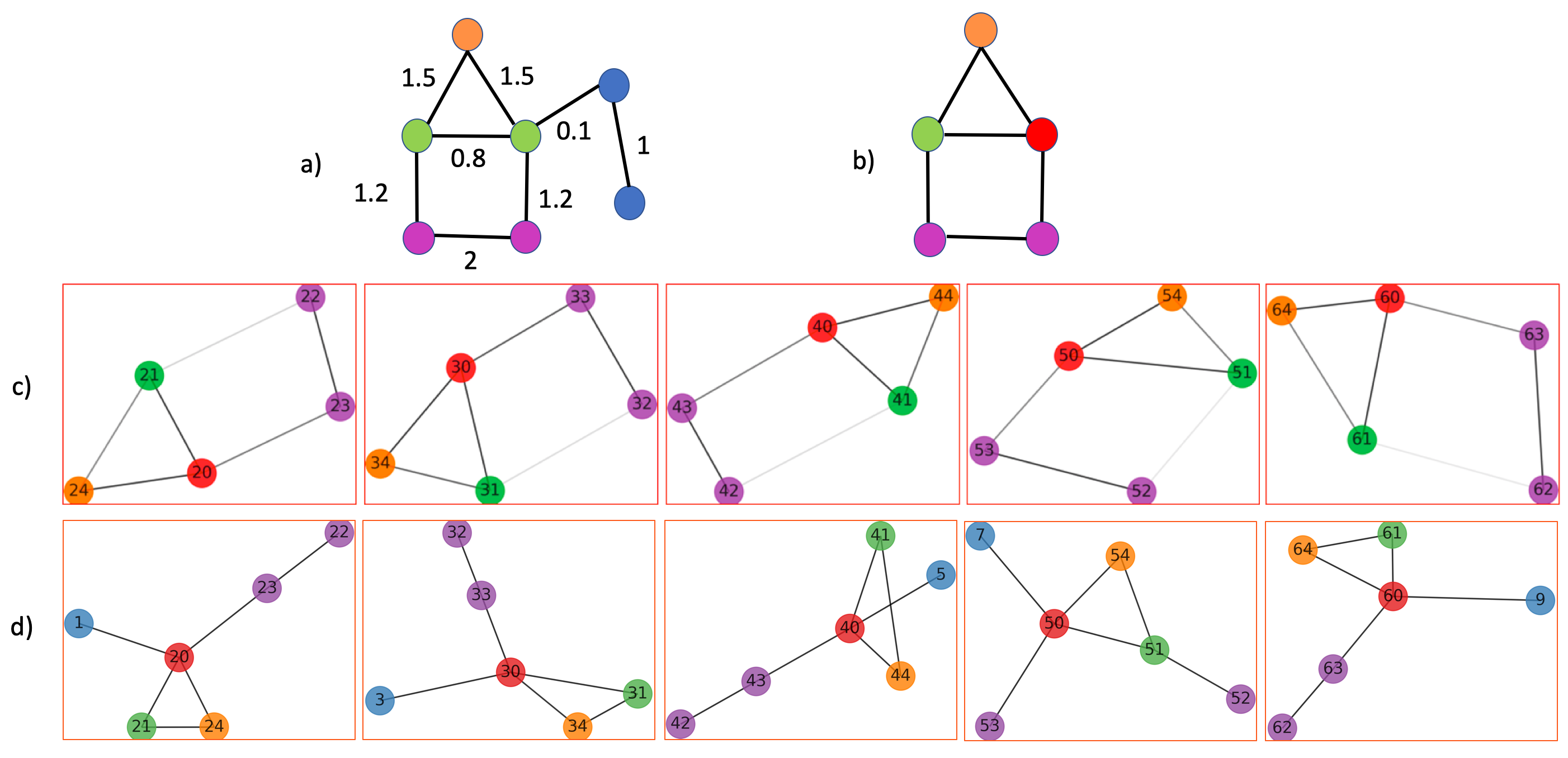}}
         	\end{minipage}
         	\caption{Comparing with GNNExplainer \cite{ying2019gnn} on SynComp dataset: a) the motif and corresponding edge weights; b) the human interpretation of the informative component to classify a node (colored in red) as class "1"; c) informative components of nodes classified as class "1" detected by our proposed MMI mask, which correctly detects the house structure; d) and informative components of nodes classified as class "1" detected by GNNExplainer \cite{ying2019gnn}, which wrongly includes the unimportant connection to BA graph. Node orders are denoted in c) and d). }
         	\label{fig:compare}
        \end{figure}
We compared our proposed MMI mask edge with  GNNExplainer \cite{ying2019gnn} for weighed graph informative components detection. Given the pretrained GNN $\Phi$, GNNExplainer learned a mask on edges and used a sigmoid function to bound each entry of the mask to $[0,1]$. Then the mask weights were used as edge weights inputted to $\Phi$. 

Remind that we created SynComp dataset by generating a Barabasi–Albert (BA) graph with 15 nodes and attaching 10 five-node house-structure graph motifs to 10 random BA nodes. Each BA node belongs to class" 0" and colored in blue. Each node on "house" belongs to class "1-3" based on its motif, and we define: the nodes on the house shoulder (colored in green) belong to class "1"; the nodes on the house bottom (colored in purple) belong to class "2"; and the node on house top (colored in orange) belong to class "3".  We performed the detection of the informative components for all the nodes on the house shoulder, which connect to a BA graph node as well. We set the connection with a small edge weights $w=0.1$ in SynComp dataset (shown in Figure \ref{fig:compare}  a) with all edge weights denoted), which meant the connection was not important compared to other edges. 

The informative components detection results are shown in Figure \ref{fig:compare} c) and d) for our proposed method and GNNExplainer correspondingly. We used human interpretation that a node on the house shoulder should belong to class "1" as ground truth. Therefore, the ground truth of the informative components to classify a node in class "1" should be a "house" structure (shown as Figure \ref{fig:compare} b), the node we try to classify is colored in red). Because no matter the node connects to a BA node or not, once it is on the "house" shoulder, it belongs to class "1". Obviously, our methods could accurately detect the 'house' structure, while directly applied GNNExplainer on weighted graph resulted in wrongly including the edge to BA nodes, as GNNExplainer ignore edge weights.
\end{document}